\documentclass[journal=ancac3,manuscript=article]{achemso}

\usepackage{graphicx}
\usepackage{dcolumn}
\usepackage{bm}
\newcommand{\nc}{\newcommand}
\nc{\be}{\begin{equation}}
\nc{\ee}{\end{equation}}
\nc{\bea}{\begin{eqnarray}}
\nc{\eea}{\end{eqnarray}}
\nc{\bean}{\begin{eqnarray*}}
\nc{\eean}{\end{eqnarray*}}
\nc{\mb}{\mbox}
\nc{\rnc}{\renewcommand}
\nc{\vk}{\mb{\bf k}}
\nc{\vK}{\mb{\bf K}}
\nc{\vp}{\mb{\bf p}}
\nc{\vn}{\mb{\bf n}}
\nc{\vq}{\mb{\bf q}}
\nc{\rr}{\mb{\bf r}}
\nc{\vz}{\hat {\mb{\bf z}}}
\nc{\vj}{\mb{\boldmath$j$}}
\nc{\vg}{\mb{\boldmath$g$}}
\nc{\x}{\mb{\boldmath$x$}}
\nc{\A}{\mb{\boldmath$A$}}
\nc{\va}{\mb{\boldmath$a$}}
\nc{\vs}{\mb{\boldmath$\sigma$}}
\nc{\vpi}{\mb{\boldmath$\pi$}}
\nc{\nab}{\nabla}
\nc{\X}{\sf x}

\title{Imaging Universal Conductance Fluctuations in Graphene}

\author{Mario F. Borunda}
\affiliation[Department of Physics, Harvard University]{Department of Physics, Harvard University, Cambridge, MA 02138, USA}
\email{mario.borunda@okstate.edu}
\author{Jesse Berezovsky}
\affiliation[Case Western Reserve University]{Department of Physics, Case Western Reserve University, Cleveland, OH 44106, USA}
\alsoaffiliation[School of Engineering and Applied Sciences, Harvard University]{School of Engineering and Applied Sciences}
\author{Robert M. Westervelt}
\affiliation[Harvard University]{Department of Physics, Harvard University, Cambridge, MA 02138, USA}
\alsoaffiliation[School of Engineering and Applied Sciences, Harvard University]{School of Engineering and Applied Sciences}
\author{Eric J. Heller}
\affiliation[Harvard University]{Department of Physics, Harvard University, Cambridge, MA 02138, USA}
\alsoaffiliation[Department of Chemistry and Chemical Biology, Harvard University]{Department of Chemistry and Chemical Biology, Harvard University, Cambridge, MA 02138, USA}

\begin{document}


\date{Apr. 5, 2011}

\begin{abstract}
We study conductance fluctuations (CF) and the sensitivity of the conductance to the motion of a single scatterer in two-dimensional massless Dirac systems. 
Our extensive numerical study finds limits to the predicted universal value of CF. 
We find that CF are suppressed for ballistic systems near the Dirac point and approach the universal value at sufficiently strong disorder.
The conductance of massless Dirac fermions is sensitive to the motion of a single scatterer.
CF of order $e^2/h$ result from the motion of a single impurity by a distance comparable to the Fermi wavelength.  
This result applies to graphene systems with a broad range of impurity strength and concentration while the dependence on the Fermi wavelength can be explored {\em via} gate voltages.
Our prediction can be tested by comparing graphene samples with varying amounts of disorder and can be used to understand interference effects in mesoscopic graphene devices.
\end{abstract}

\maketitle
\noindent{\bf keywords:} graphene $\cdot$ transport property $\cdot$ conductance fluctuations $\cdot$ electronic transport in nanoscale materials and structures $\cdot$ nanoelectronic devices 

The two-dimensional (2D) Dirac equation is relevant in elucidating the electronic and transport properties of recently discovered materials such as graphene and topological insulators.
Graphene~\cite{Geim:2007,castro_rpm}, a one-atom-thick allotrope of carbon, and the topological surface states of materials with a bulk gap such as Bi$_{1-x}$Sb$_x$ crystals~\cite{topo_intro} share unusual Dirac-like electronic structure providing an enthralling test bed for new physics and inevitable future applications based on quantum interference effects~\cite{Berger_2006,Heersche_2007, Lundeberg_2009}. 
Given the Dirac nature of the electronic spectrum, the quasiparticles in graphene propagate as massless relativistic particles, making graphene a qualitatively different material from conventional conductors.
At low temperatures, the quantum transport of electrons becomes coherent and leads to quantum interference corrections to the conductance. The amplitude of these corrections follows a universal scaling with value $\delta G \approx e^2/h$ receiving the name of universal conductance fluctuations (UCF)~\cite{Altshuler:1985,Lee:1985}. 
The conductance fluctuations (CF) take place when a coherent electron wave scatters repeatedly while traversing a disordered conductor. 
The wave follows all possible paths through the sample and different paths interfere with each other, giving rise to CF that are independent of the sample size and the degree of disorder. 
Due to the CF, the conductivity is sensitive to changes in the configuration of the impurity scatterers.
Theoretical studies~\cite{Altshuler:1985c, Feng:1986} have predicted that the full UCF effect is obtained by moving a single scatterer a distance comparable to half the Fermi wavelength $\lambda_F$. 
In gate-doped graphene the gate voltage controls the average charge density of the device and given the relationship $\lambda_F = 2 (\pi/ \vert n \vert)^{0.5}$, the gate also presents an exquisite control over the Fermi wavelength.
The Fermi wavelength diverges as the carrier density approaches the neutrality (Dirac) point, making the low-density limit particularly interesting given that experimentally the conductance does not go to zero and disagreements between theory and experiments regarding the minimum value of the conductivity exist. 

Recent experimental and theoretical studies of the CF in graphene have been reported~\cite{Berger_2006,Heersche_2007, Lundeberg_2009,Rycerz:2007,Cheianov:2007, Staley:2008, Kharitonov:2008,Kechedzhi:2008,Kechedzhi:2009,Chen:2010,Ojeda:2010,Berezovsky:2009,Berezovsky:2009b}.
Berezovsky and colleagues~\cite{Berezovsky:2009,Berezovsky:2009b} demonstrated unambiguously the sensitivity of the CF to the motion of a single scatterer. 
Placing a charged scanning probe microscope tip~\cite{Topinka:2003} near the graphene sample creates an image charge that acts as a movable scatterer. 
By mapping the CF {\em versus} scatterer position, it was found that the UCF decorrelate when the induced scatterer is displaced by a distance comparable to half the Fermi wavelength~\cite{Berezovsky:2009}.
  
In this article we calculate the zero temperature CF for quasiparticles obeying the massless 2D Dirac equation in the presence of disorder, assuming elastic scattering by fixed scatterers and with no sources of inelastic scattering. 
The motivation for the study of this system is threefold: First, the massless Dirac Hamiltonian model is of relevance to transport in graphene and the surface states of topological insulators.
Second, we probe the sensitivity of the conductance to changes in the impurity strengths over a range of carrier densities and system sizes that are accessible by current experiments. 
Finally, we calculate the CF due to the motion of a single scatter.
This last result demonstrates the sensitivity of CF to the motion of an ionized charged impurity and shows a dependence on the carrier density that applies to all but the most disordered graphene devices. 

\section{Results and Discussion}
\textbf{Model}.
We study the following 2D Hamiltonian:
\be
H = - i \hbar v \left(  {\bm \sigma}_x \partial_x + {\bm \sigma}_y \partial_y \right)  +  V(\rr){\bm \sigma}_0
\label{ham}
\ee
with $v$ being the velocity of the Dirac fermions, ${\bm \sigma}_i$ the $2\times2$ Pauli matrices, and $V(\rr)$ a (pseudo)spin-independent potential. 
The spinor wave function applies to graphene near energies close to the Dirac (neutrality) points with each component pertaining to one of the two atoms in the unit cell.
Our study is based on a transfer matrix approach to calculate the conductance of Dirac fermions\cite{Tworzydlo_2008}, described in the methods section.
The conductance is modeled by calculating the reflection $r$ and transmission $t$ matrices in realistic devices. 
Although the use of the Dirac equation does not take into account the anisotropy of the Dirac cones (trigonal warping) in graphene, the resulting trigonal distortion occurs at doping values considerably larger than those presented here. 

Experiments have observed an electrostatic landscape with naturally occurring variations in the carrier density~\cite{Martin:2008, Zhang:2009,Deshpande:2009,Teague:2009}. 
In this study we consider such a potential arising from the random charged impurities located either in the substrate or above the graphene plane~\cite{Polini_2008,rossi:2008}. 
This type of disorder profile is both smooth and slowly varying on the atomic scale, suppressing intervalley scattering.
The disorder model used in our calculations neglects scattering from short-range atomic defects and ripples.
This approach is justified by the long-range (compared to the atomic lattice) nature of the Coulomb-like charged impurities.

After choosing the average charge density $n_0$ we introduce local variations in the charge density either through randomness of the on-site carrier density at each lattice point $\eta_j$ or from ionized impurities near the graphene plane.
In the first case, the contribution of the spatially varying on-site carrier density is distributed uniformly with width $\Delta n$ [$n(\rr) = n(r_j) = n_{0} + \eta_j$ with $-\Delta n \leq \eta_j  \leq\Delta n]$.
In the latter procedure, a fraction $n_i$ of the lattice sites are randomly chosen and an ionized impurity center is located a distance away. 
We solve for the charge density induced from this set of ideally screened Coulombic scatterers. 
Using the relationship between the Fermi energy and carrier density we obtain the local scattering potential that is used in \ref{ham}, $V(r_j) = \hbar v ~ \mbox{sgn}(n(r_j)) \sqrt{\pi n(r_j)}$.
Note that the square root means that the contributions to the potential do not add arithmetically. 
In this simple model the random on-site component dominates at low carrier density and its contribution diminishes as the overall charge density increases.
Although our approach overestimates screening at the interface of hole-rich/electron-rich regions, the charge of the scatterers is chosen to yield an rms charge density in agreement with the observed charged puddles.~\cite{Martin:2008, Zhang:2009}
A more rigorous (quantum) treatment~\cite{Polini_2008} of the density response is limited to system sizes significantly smaller than those we have investigated in the present work. 

\textbf{Conductance}.
The conductance is obtained from the Landauer formula,
\be
G = G_0 \sum_n^N T_n = G_0 \mbox{Tr}\left[t t^\dag\right]
\label{Landauer}
\ee 
where $G_0$ is the quantum of conductance ($4e^2/h$ for graphene due to spin and valley degeneracy), $N$ is the total number of transverse modes in the sample, and $T_n$ are the transmission eigenvalues obtained from the diagonalization of the matrix product $t t^\dag$. The transmission matrix $t$ is obtained using the transfer matrix approach. The details of the calculation are described in the methods section.  

\begin{figure}[!t]
\begin{center}
\includegraphics[width=0.65\textwidth]{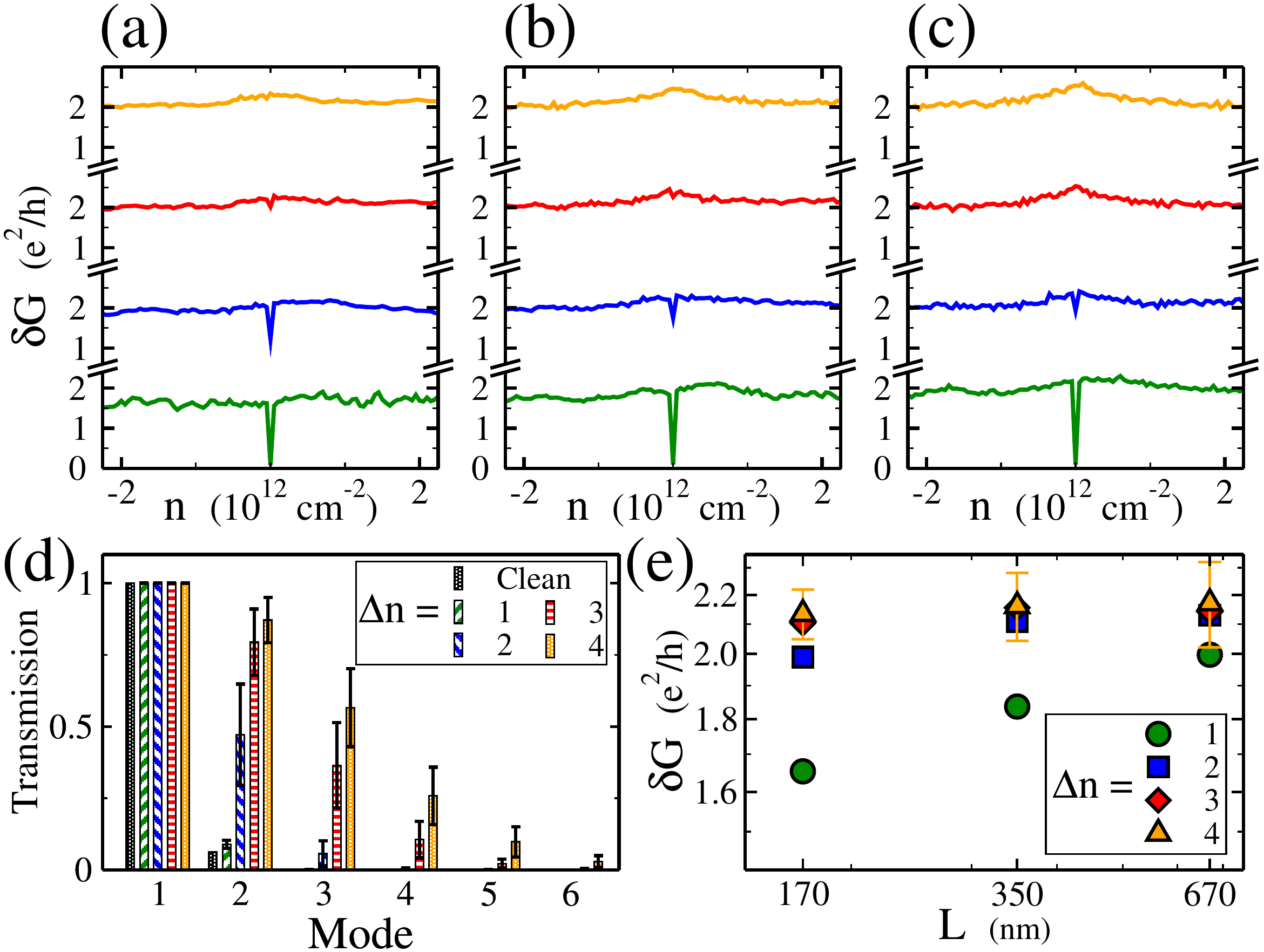}
\caption{
Conductance fluctuations of Dirac fermions. The CF are plotted in terms of the standard deviation of $G$ ($\delta G$ in $e^2/h$) versus carrier density $n$ (in cm$^{-2}$) for several values of the fluctuations in the carrier density width $\Delta n$ (from bottom to top $\Delta n= 1.0, 2.0, 3.0$, and $4.0 \times 10^{12} cm^{-2}$) and system sizes (a) $L = 170$, (b) $L = 350$, and (c) $L = 670~ nm$. (d) Transmission {\em per} mode at the Dirac point for different values of $\Delta n$. After the first mode, all other modes come in degenerate pairs with only one shown in the plot.
(e) Scaling of the CF with system size for different values of $\Delta n$. The length of the sample is varied with a fixed W/L ratio of 3.}
\label{f:varGvsn}
\end{center}
\end{figure}

\textbf{Conductance fluctuations}.
\ref{f:varGvsn} (a-c) presents the CF, with $\delta G$ defined as the standard deviation of the conductance, as a function of carrier density in the no decoherence limit. 
Measurements of the magnitude of the CF at the Dirac point have yielded mixed results. In weakly disordered bilayer and trilayer graphene devices the UCF are suppressed near the Dirac point~\cite{Staley:2008}. 
Although the presence of more than one layer of graphene substantially changes the electronic properties of the device,\cite{castro_rpm} possible explanations for the suppression of the fluctuations\cite{Staley:2008} involved a different mechanism for quantum interference of edge states, which dominate the conductance of the low-density samples. 
A reduction of the amplitude of the fluctuations was seen in monolayer graphene devices\cite{Chen:2010, Berezovsky:2009} contrasting with a study\cite{Ojeda:2010} in which the amplitude of the fluctuations is larger at the charge neutrality point for both monolayer and bilayer graphene devices. 
Our calculations, covering a broad range of impurity strengths and concentrations, finds that the amplitude of the conductance fluctuations at the Dirac point for sufficiently strong fluctuations in the carrier density $\Delta n >  2.0\times 10^{12} cm^{-2}$ exhibits a peak. 
In contrast, the CF decrease at exactly the Dirac point for weak on-site density fluctuations. 

The suppression of the CF in weakly disordered systems is due to the reduced size of the fluctuations in the cleaner systems coupled to a reduced number of modes open to conduct~\cite{footnote2}.  
\ref{f:varGvsn} (d) shows the transmission for the first modes at the neutrality point for several values of the on-site carrier density averaged over several disorder configurations in $L=350~nm$ systems.
All systems, independent of the amount of disorder present, have at one mode completely open.
In the clean systems, the second mode is only partially open ($T=0.06$).
Disorder opens up the subsequent modes for transport as can be seen in \ref{f:varGvsn} (d) where error bars show the fluctuation for each of the modes.
Assuming that the CF are proportional to the fluctuations in each of the modes, the suppression of the CF is due to the smaller fluctuations in the cleaner systems coupled to the reduced number of modes open to transport.   

In \ref{f:varGvsn} (e) we plot the averaged CF as a function of system size. 
For systems with a fixed W/L ratio of 3, we vary the length from $170~nm$ to $670~nm$. 
As the strength of the on-site carrier density increases, the value of $\delta G$ converges to $2.17\pm0.15 $ $e^2/h$, close to the value found analytically for doped graphene of 2.36 $e^2/h$ in the W/L $\gg$ 1 aspect ratio limit~\cite{Kechedzhi:2008,Kharitonov:2008}.
The value is higher than the UCF value ($e^2/h$) due to the absence of intervalley scattering and trigonal warping, both absent in our model.

\begin{figure}[!t]
\begin{center}
\includegraphics[width=0.65\textwidth]{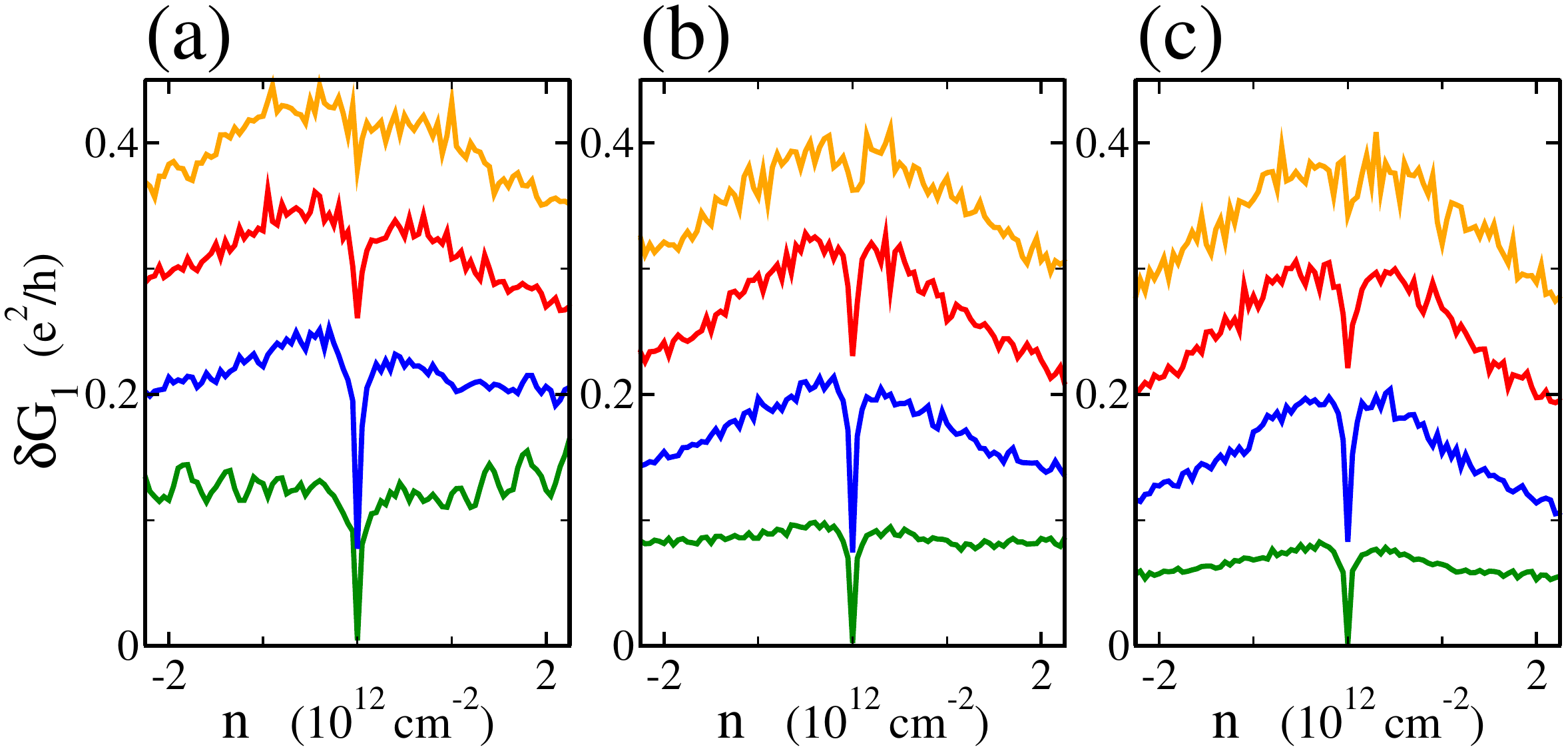}
\caption{
Conductance fluctuations due to the interchange of two on-site densities.
The rms deviation of the conductance upon interchange of the local charge density of two sites ($\eta_j$) for several values of the fluctuations in the carrier density with width $\Delta n$ (from bottom to top $\Delta n = 1.0, 2.0, 3.0$, and $4.0 \times 10^{12} cm^{-2}$) and system sizes (a) $L = 170$, (b) $L = 350$, and (c) $L = 670~ nm$.}
\label{f:deltaG1}
\end{center}
\end{figure}

\textbf{Sensitivity to motion of a single impurity}.
Similarly, we compare the CF from different impurity configurations to the CF induced by the motion of a single impurity.
The method developed by Tworzydlo {\em et al.}~\cite{Tworzydlo_2008} is suitable to revisit a landmark study carried out by Feng {\em et al.} for metallic systems~\cite{Feng:1986}.
We obtain the conductance of massless Dirac fermions for a discrete lattice model with random on-site charge densities (similar to the Anderson model) and calculate the rms deviation of this conductance $\delta G_1$ after the interchange of on-site charge densities of just one pair of sites.
In \ref{f:deltaG1} we plot $\delta G_1$ as a function of carrier density for several system sizes averaged over different realizations of the random densities and for several widths $\Delta n$ over which these on-site values are chosen. 
While an increase in the strength of the on-site carrier density produces an increase in the rms value of CF, we found $\delta G_1$ is independent of sample size.
Our calculation chooses two sites to interchange at random and as such their separation can be small or their on-site values similar.
Thus, although the average value of $\delta G_1$ is lower than the average value of $\delta G$, in general $\delta G_1$ is bounded by $\delta G$. 

For weak disorder and small system size ($\Delta n= 1.0 \times 10^{12}~cm^{-2}$ and $L = 170~nm$), CF for both the complete configuration change (lower curve in \ref{f:varGvsn} (a)) and upon interchange of two sites (lower curve in \ref{f:deltaG1} (a)) exhibit oscillations as a function of the carrier density. 
In such systems ballistic transport dominates, {\em i.e.}, the transport mean free path is larger than the system size $l > L$. 
The oscillations in the CF as a function of density are caused by multiple reflections at the ends of the sample (Fabry-Perot resonances) where the enhancements are due to multiple visits~\cite{Liang:2001,Du:2008}.
 
\begin{figure}[!t]
\begin{center}
\includegraphics[width=0.65\textwidth]{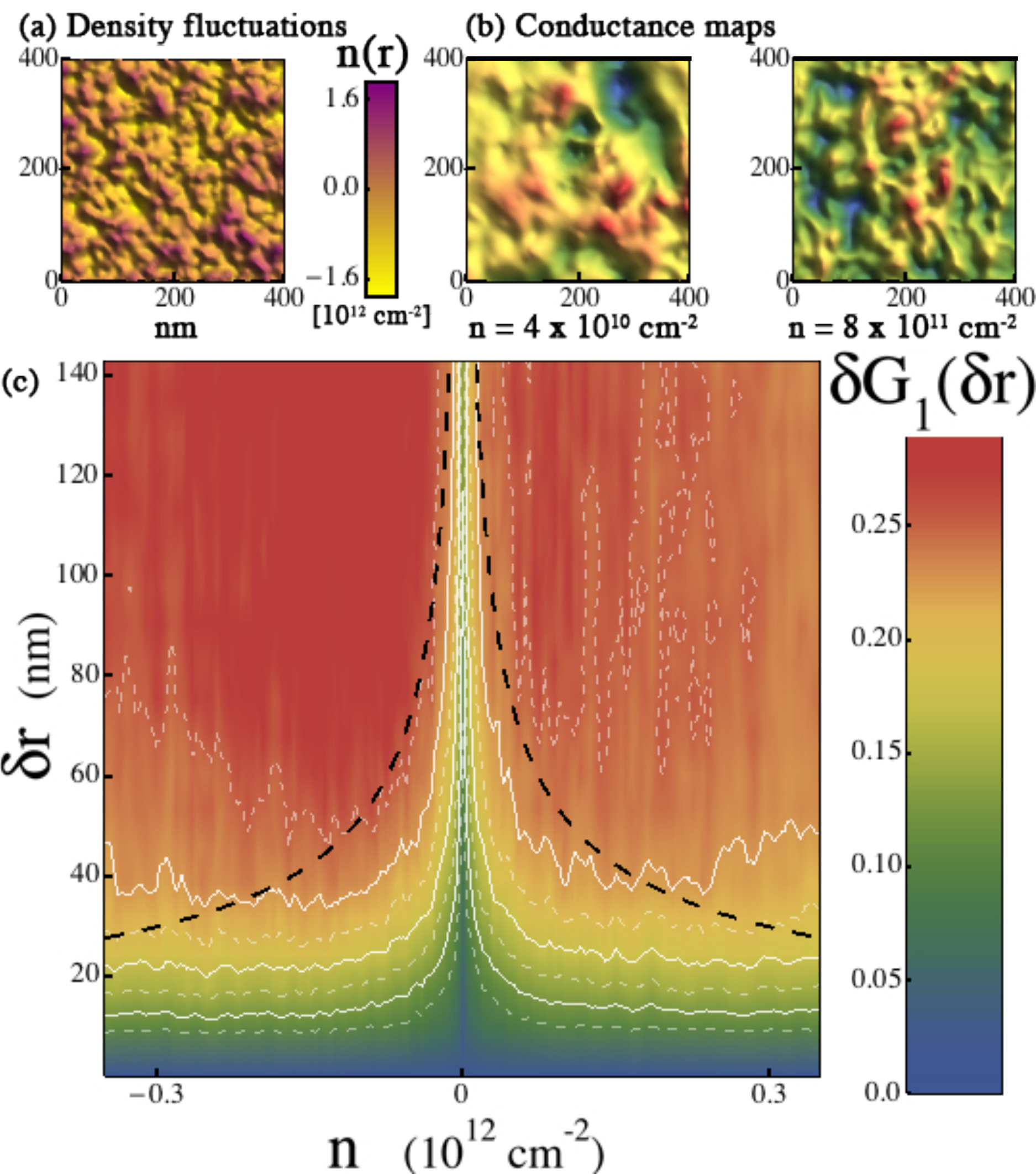}
\caption{
Evidence of the conductance sensitivity of Dirac fermions. 
(a) Typical carrier density fluctuations resulting from random charged impurities localized in the substrate or above the plane in a graphene device. 
The  concentration of charged impurities is 20 \% of sites, and are located 1 to 10 $nm$ away from the graphene plane with a charge of $\pm 2$ $e$, yielding density fluctuations $n(r) \sim 1.6 \times 10^{12} cm^{-2}$ for system size W = 1.005 $\mu m$ and L = 670 $nm$.  
(b) Conductance maps, the device conductance versus the position of the movable scatterer, at two different carrier densities simulated by raster scanning a single charged impurity over a 400 x 400 $nm^2$ area of the sample. 
The maps display spatial CF $\delta G_1 \sim e^2/h$ with the lateral size of the fluctuations features depending on the carrier density.
(c) Plot of the conductance change $\delta G_1$ {\em versus} the charge density $n$ and the distance $\delta r$ that a single ionized impurity moved, averaged over 2000 impurity ensembles.
The black dashed line is presented as a guide indicating the length $l_r$ where UCF theory predicts that the motion of a single impurity will induce a change in the CF comparable to that of a new configuration of impurities.}
\label{f:deltaG1(r)}
\end{center}
\end{figure}

Finally, we consider the change in con\-duct\-ance induced by the motion of a single charged impurity a distance $\delta r$.
A fraction of lattice sites are ran\-dom\-ly selected and charged impurities are placed above their positions ($r_1, r_2, ..., r_{N_i}$) inducing a charge density landscape such as the one presented in \ref{f:deltaG1(r)} (a). 
Such an experiment was undertaken by Berezovsky {\em et al.} where the device conductance was measured as a function of SPM tip position, and the movable scatterer is created by the SPM tip~\cite{Berezovsky:2009}. 
We present such conductance maps obtained from raster scanning one of the impurities in \ref{f:deltaG1(r)} (b).
Maps are presented for two different carrier densities and show that the lateral size of the fluctuations depends on the carrier density.
It was shown that the length $l_r$ that a strong scattering center needs to be shifted to decorrelate the CF ({\em i.e.,} to change the conductance by $\sim e^2/h$) is approximately half the Fermi wavelength~\cite{Berezovsky:2009}. 
This length was obtained from the autocorrelation function of the conductance maps and studied as a function of the carrier density~\cite{Berezovsky:2009}.
In this new study, we calculate the conductance change for the motion of one impurity, 
\be
\delta G_1(\delta r) = \sqrt{\left( G(r_1, ..., r_{N_i}) - G(r_1, ..., r_{N_i}+\delta r) \right)^2},
\label{conductance_change}
\ee
over a range of system sizes and carrier densities for several impurity densities, accessible to current experiments. 

For diffusive metallic systems $\delta G_1$ is given by~\cite{Feng:1986}
\be
\left(\delta G_1\right)^2 \approx 
\frac{e^4}{h^2} \frac{\Omega}{N_i l^d} \left(\frac{L}{l} \right)^{2-d}
\left[ 1 - \left( \frac{\sin \frac{1}{2} k_F \delta r}{\frac{1}{2}k_F \delta r} \right)^2 \right]
\label{conductance_change_T}
\ee 
where $\Omega$ is the volume and $d$ the dimensionality.
In 2D systems, the $L/l$ factor drops out and $\Omega/N_i l^2$ corresponds to the strength of the impurities. 
\ref{conductance_change_T} predicts that if $k_F \delta r \geq 1$ the change in conductance from the motion of a single impurity will be comparable to the complete change of the impurity configuration in a sample.  
In \ref{f:deltaG1(r)} (c) we show the rms fluctuations of the conductance as a function of carrier density and distance moved by the impurity.
The dashed line in the plot indicates the length $l_r = 0.46 \lambda_F$ separating the regions at which motion of a single impurity will impact the CF as if the impurity configuration was completely different.
As evident in \ref{f:deltaG1(r)} (c), these computed results indicate a deviation from conventional UCF theory~\cite{Feng:1986} for 2D electron gas of Dirac fermions. 
In connection with previous studies~\cite{Berezovsky:2009}, the $l_r$ length obtained from the autocorrelation function will saturate at certain values of the carrier density.
One of the main findings of this letter is that the trend presented in \ref{f:deltaG1(r)} (c) applies to samples where the carrier-density fluctuations ranges from $10^9$ to $3 \times 10^{12} cm^{-2}$, {\em i.e.,} for all but the dirtiest graphene devices.

\section{Conclusion}
In conclusion, we have focused on the CF of Dirac fermions and the impact of the motion of a single impurity.  
Our results predict that the CF in 2D Dirac systems is dependent on the strength of the disorder near the neutrality point in ballistic systems but independent of the strength of the disorder for doped graphene. 
For strong enough disorder, as is the case of graphene on SiO$_2$ substrates, the CF will not depend strongly on carrier density but are enhanced at the Dirac point.
Studying CF on suspended graphene~\cite{Du:2008} and graphene deposited on hexagonal boron nitride substrates~\cite{Dean:2010} can test these predictions. 
Consistent with theoretical predictions, the change in conductance caused by the motion of a single impurity (covering a small area of the sample) is significant when the distance moved is of the order of the Fermi wavelength.   

\section{Methods}
\textbf{Transfer matrix approach}.
We sketch the transfer matrix procedure used, for further details we refer to Tworzydlo {\em et al.}~\cite{Tworzydlo_2008}. 
The system is discretized into a lattice and the difference equations are solved without violating symplectic symmetry and current conservation while avoiding the fermion doubling problem~\cite{Tworzydlo_2008}.
We calculate the conductance in a strip geometry, discretizing the sample using a square lattice. 
The longitudinal direction extends from $x = 0$ to $x = L$ and the transverse direction from $y = 0$ to $y = W$, where $L$ and $W$ are the length and width of the sample. 
Periodic boundary conditions are used in the transverse direction. 
The transfer matrix reads
\be
\Psi_{m+1} = {\cal M}_m \Psi_m
\label{TransferMat}
\ee
where $\Psi_m$ is a vector containing the values for the wavefunction $\Psi(x,y)$ at $x= m a_0$, with $m$ being an integer and $a_0 = 10~nm$ the lattice spacing.  
Semi-infinite metallic leads are attached to the strip at its ends ($x=0$ and $x=L$).
The metal contacts are ballistic leads in which all modes are conducting~\cite{footnote1}. Each incoming mode on a lead is propagated to the other lead using the transfer matrix.
The $N$ transverse modes in the sample are either propagating modes $\phi_l$ or evanescent modes $\chi_l$ (modes that decay for large positive or negative values of $x$). 
An incoming wavefunction in mode $l_0$ starting on the left side of the sample is composed of incoming, reflected, and evanescent modes 
\be
\Phi_{l_0}(x=0) = \phi^+_{l+0} + \sum_l r_{l,l_0} \phi_l^- + \sum_l \alpha_{l,l_0} \chi_l^-
\ee
at the $x=L$ edge of the sample the wavefunction is the sum of the transmitted and evanescent waves given by
\be
\Phi_{l_0}(x=L) = \sum_l t_{l,l_0} \phi_l^+ + \sum_l \alpha'_{l,l_0} \chi_l^+
\ee
where the label $+$ corresponds to right moving and $-$ to left moving modes.
The reflection $r_{l,l_0}$ and transmission coefficient $t_{l,l_0}$ are obtained from the transfer matrix relation
\be
\Phi_{l_0}(x=L) = {\cal M}  \Phi_{l_0}(x=0)
\label{TransferMat2}
\ee
and elimination of the $\alpha$ and $\alpha'$ coefficients.
Once this is done for all possible modes the reflection $r$ and transmission $t$ matrices are composed. 
Similarly, repeating the procedure for modes propagating from the right to the left edge of the sample results in the $r'$ and $t'$ matrices.\cite{Tworzydlo_2008}
The conductance is calculated using the Landauer formula, \ref{Landauer}, by summing over the transmission eigenvalues $T_n$ obtained from the diagonalization of the matrix product $t t^\dag$.

\textbf{Acknowledgements}. Discussions with M.~C. Barr, H. Hennig, A. Jurisch, and Y. Vasquez are gratefully acknowledged. 
The calculations in this paper were run on the Odyssey cluster supported by the FAS Research Computing Group at Harvard University.  
Research was supported by the U.S. Department of Energy, Office of Basic Energy Sciences, M.F.B. and E.J.H. by DOE BES DE-FG02-08ER46513 and J.B. and R.M.W. by DOE BES DE-FG02-07ER46422.


\begin{thebibliography}{99}

\bibitem{Geim:2007} 
Geim, A.~K.; Novoselov, K. S.
The Rise of Graphene.
{\em Nature Mat.}  {\bf 2007}, {\em 6}, 183-191.
\bibitem{castro_rpm} 
Castro~Neto, A.~H.; Guinea, F.; Peres, N.~M.~R.; Novoselov, K. S.; Geim, A.~K. 
The Electronic Properties of Graphene. {\em Rev. Mod. Phys.} {\bf 2009}, {\em 81}, 109-162.

\bibitem{topo_intro}
Hasan, M.~Z.; Kane, C.~L. 
Colloquium: Topological Insulators.
{\em Rev. Mod. Phys.} {\bf 2010}, {\em 82}, 3045-3067.

\bibitem{Berger_2006}
Berger, C.; Song, Z.; Li, X.; Wu, X.; Brown, N.; Naud, C.; Mayou, D.; Li, T.; Hass, J.; Marchenkov, A. N.; Conrad, E. H.; First, P. N.; de Heer, W. A.
Electronic Confinement and Coherence in Patterned Epitaxial Graphene. 
{\em Science} {\bf 2006}, {\em 312}, 1191-1196.
\bibitem{Heersche_2007}
Heersche, H.~B.; Jarillo-Herrero, P.; Oostinga, J.~B.; Vandersypen, L.~M.~K.; Morpurgo, A.~F.;
Bipolar Supercurrent in Graphene.
{\em Nature} {\bf 2007}, {\em 446}, 56-59.
\bibitem{Lundeberg_2009}
Lundeberg, M.~B.; Folk, J.~A.
Spin-resolved Quantum Interference in Graphene.
{\em Nature Phys.} {\bf 2009}, {\em 5}, 894-897. 

\bibitem{Altshuler:1985} 
Altshuler, B.~L.
Fluctuations in the Extrinsic Conductivity of Disordered Conductors. 
{\em Pis'ma Zh. Eksp. Teor. Fiz.} {\bf 1985}, \textit{41}, 530-533 [{\em JETP Lett.} {\bf 1985}, {\em 41}, 648-651].

\bibitem{Lee:1985}
Lee, P.~A.; Stone, A.~D.
Universal Conductance Fluctuations in Metals.
{\em Phys. Rev. Lett.} {\bf 1985}, {\em 55}, 1622-1625.

\bibitem{Altshuler:1985c} Altshuler, B.~L.; Spivak, B.~Z.
Variation of the Random Potential and the Conductivity of Samples of Small Dimensions.
{\em Pis'ma Zh. Eksp. Teor. Fiz.} {\bf 1985}, {\em 42}, 363-365 [{\em JETP Lett.} {\bf 1985}, {\em 42}, 447-450].

\bibitem{Feng:1986} Feng, S.; Lee, P.~A.; Stone, A.~D.
Sensitivity to the Conductance of a Disordered Metal to the Motion of a Single Atom: Implication for 1/\textit{f} Noise.
{\em Phys. Rev. Lett.} {\bf 1986}, {\em 56}, 1960-1963.

\bibitem{Rycerz:2007} Rycerz A.; Tworzydlo, J.; Beenakker, C.~W.~J.
Anomalously Large Conductance Fluctuations in Weakly Disordered Graphene.
{\em Europhys. Lett.} {\bf 2007}, {\em 79}, 57003.

\bibitem{Cheianov:2007} Cheianov, V.~V.; Falko, V.~I.; Altshuler, B.~L.; Aleiner, I.~ L.
Random Resistor Network Model of Minimal Conductivity in Graphene. 
{\em Phys. Rev. Lett.} {\bf 2007}, {\em 99}, 176801.

\bibitem{Staley:2008} Staley, N.~E.; Puls, C.~P.; Liu, Y. 
Suppression of conductance fluctuation in weakly disordered mesoscopic graphene samples near the charge neutral point.
{\em Phys. Rev. B} {\bf 2008}, {\em 77}, 155429. 

\bibitem{Kechedzhi:2008} Kechedzhi, K.; Kashuba, O.; Falko, V.~I.
Quantum kinetic equation and universal conductance fluctuations in graphene. 
{\em Phys. Rev. B} {\bf 2008}, {\em 77}, 193403.

\bibitem{Kharitonov:2008} Kharitonov, M.~Y.; Efetov, K.~B.
Universal conductance fluctuations in graphene. 
{\em Phys. Rev. B} {\bf 2008}, {\em 78}, 033404.

\bibitem{Kechedzhi:2009} Kechedzhi, K.; Horsell, D.~W.; Tikhonenko, F.~V.; Savchenko, A.~K.; Gorbachev, R.~V.; Lerner, I.~V.; Falko, V.~I.
Quantum Transport Thermometry for Electrons in Graphene. 
{\em Phys. Rev. Lett.} {\bf 2009}, {\em 102}, 066801.

\bibitem{Chen:2010} Chen, Y.-F.; Bae, M.-H.; Chialvo, C.; Dirks, T.; Bezryadin, A.; Mason, N.
Magnetoresistance in single-layer graphene: weak localization and universal conductance fluctuation studies.
{\em J. Phys.: Condens. Matter} {\bf 2010}, {\em 22}, 205301.

\bibitem{Ojeda:2010} Ojeda-Aristizabal, C.; Monteverde, M.; Weil, R.; Ferrier, M.; Gueron, S.; Bouchiat, H.
Conductance Fluctuations and Field Asymmetry of Rectification in Graphene.
{\em Phys. Rev. Lett.} {\bf 2010}, {\em 104}, 186802.

\bibitem{Berezovsky:2009} Berezovsky, J.; Borunda,  M.~F.; Heller, E.~J.; Westervelt, R.~M.
Imaging Coherent Transport in Graphene (Part I): Mapping Universal Conductance Fluctuations.
{\em Nanotechnology} {\bf 2010}, {\em 21}, 274013.

\bibitem{Berezovsky:2009b}  Berezovsky, J.; Westervelt, R.~M.
Imaging Coherent Transport in Graphene (Part II): Probing Weak-Localization.
{\em Nanotechnology} {\bf 2010}, {\em 21}, 274014.

\bibitem{Topinka:2003} Topinka, M.~A.; Westervelt, R.~M.; Heller, E.~J.
Imaging Electron Flow.
{\em Phys. Today} {\bf 2003}, {\em 56}, 47.

\bibitem{Tworzydlo_2008} Tworzydlo, J.; Groth, C. W.; Beenakker, C. W. J. 
Finite difference method for transport properties of massless Dirac fermions.
{\em Phys. Rev. B} {\bf 2008}, {\em 78}, 235438.

\bibitem{Martin:2008} 
Martin, J.; Akerman, N.; Ulbricht, G.; Lohmann, T.; Smet, J.~H.; von Klitzing, K.; Yacoby, A.
Observation of Electron-Hole Puddles in Graphene Using a Scanning Single-Electron Transistor.
{\em Nature Phys.} {\bf 2008}, {\em 4}, 144-148.

\bibitem{Zhang:2009} 
Zhang, Y.; Brar, V.~W.; Girit, C.; Zettl, A.; Crommie, M.~F.
Origin of Spatial Charge Inhomogeneity in Graphene.
{\em Nature Phys.} {\bf 2009}, {\em 5}, 722-726.
\bibitem{Deshpande:2009} 
Deshpande, A.; Bao, W.; Miao, F.; Lau, C.~N.; LeRoy, B.~J.
Spatially resolved spectroscopy of monolayer graphene on SiO$_2$.
{\em Phys. Rev. B} {\bf 2009}, {\em 79}, 205411.
\bibitem{Teague:2009} 
Teague, M.~L.; Lai, A.~P.; Velasco, J.; Hughes, C.~R.; Beyer, A.~D.; Bockrath, M.~W.; Lau, C.~N.; Yeh, N.-C.
Evidence for Strain-Induced Local Conductance Modulations in Single-Layer Graphene on SiO$_2$.
{\em Nano Lett.} {\bf 2009}, {\em 9}, 2542-2546.

\bibitem{Polini_2008} Polini, M.; Tomadin, A.; Asgari, R.; MacDonald, A. H. 
Density functional theory of graphene sheets.
{\em Phys. Rev. B} {\bf 2008}, {\em 78}, 115426.

\bibitem{rossi:2008} Rossi, E.; Das Sarma, S.
Ground State of Graphene in the Presence of Random Charged Impurities.
{\em Phys. Rev. Lett.} {\bf 2008}, {\em 101}, 166803.

\bibitem{footnote1} As explained by Tworzydlo {\em et al.}\cite{Tworzydlo_2008}, the large Fermi wave vector in the leads excite spurious evanescent modes in the sample. Due to an artifact of the discretization procedure, along with purely imaginary evanescent modes ($k = i \kappa_+$), complex modes appear ($k = \pi + i \kappa_-$). Between the leads and the sample there is a region to filter such modes.

\bibitem{footnote2} The symplectic symmetry of the Dirac Hamiltonian (\ref{ham}) requires the transfer matrix (\ref{TransferMat}) to obey certain conditions such as ${\cal M}_m = \sigma_y {\cal M}_m^* {\cal M}_m$ \cite{Tworzydlo_2008}.
As a result, the transmission eigenvalues used in the Landauer formula (\ref{Landauer}), which are always an odd number due to the boundary conditions, are composed of one mode that is always open ($T=1$) followed by the rest of the modes, which come as sets of degenerate pairs (due to Kramers degeneracy).

\bibitem{Liang:2001}
Liang, W; Bockrath, M.; Bozovic, D.; Hafner, J. H.; Tinkham, M.; Park, H.
Fabry - Perot interference in a nanotube electron waveguide.
{\em Nature} {\bf 2001}, {\em 411}, 665-669.


\bibitem{Du:2008} Du, X.; Skachko, I.; Barker, A.; Andrei, E.~Y.
Approaching ballistic transport in suspended graphene.
{\em Nature Nanotech.} {\bf 2008}, {\em 3}, 491-495.

\bibitem{Dean:2010} Dean, C.~R.; Young, A.~F.; Meric, I.; Lee, C.; Wang, L.; Sorgenfrei, S.; Watanabe, K.; Taniguchi, T.; Kim, P.; Shepard, K. L.; Hone, J.
Boron Nitride Substrates for High-Quality Graphene Electronics.
{\em Nature Nanotech.} {\bf 2010}, {\em 5}, 722-726.

\end{thebibliography}
\end{document}